\documentclass[fleqn,12pt,twoside]{article}
\usepackage[headings]{espcrc1}

\readRCS
$Id: bzhang.tex,v 1.5 2005/09/29 21:30:56 bzhang Exp bzhang $
\usepackage{graphicx}
\usepackage[figuresright]{rotating}

\newcommand{\AmS}{{\protect\the\textfont2
  A\kern-.1667em\lower.5ex\hbox{M}\kern-.125emS}}

\title{Charm elliptic flow in Au+Au collisions at RHIC
        \thanks{
        Work supported by the U.S. National Science Foundation 
        under Grant No.'s PHY-0140046 (BZ) and PHY-0457265 (CMK),
	the Welch Foundation under Grant No. A-1358 (CMK), and the 
	National Natural Science Foundation of China under Grant 
	No.'s 10105008 and 10575071(LWC).
}
}

\author{Bin Zhang\address[ASU]{Department of Chemistry and Physics,
        Arkansas State University,
        P.O. Box 419, State University, AR 72467-0419, USA},
        Lie-Wen Chen\address[SJTU]{Institute of Theoretical Physics,
        Shanghai Jiao Tong University, Shanghai 200240, China}
        and
        Che-Ming Ko\address[TAMU]{Cyclotron Institute and Physics Department,
        Texas A\&M University, College Station, Texas 77843-3366, USA}
}
       
\runtitle{Charm elliptic flow at RHIC}
\runauthor{B. Zhang \textit{et. al}}

\begin{document}

\maketitle

\begin{abstract}

Using the perturbative method for the simulation of charmed particles, 
the dynamical origin of charm quark elliptic flow is studied in the 
framework of a multi-phase transport (AMPT) model.  Besides the expected
ordering relative to that of light quarks according to quark masses, 
charm quark elliptic flow is seen to be sensitive to the parton 
scattering cross section.  To describe the observed large elliptic 
flow of electrons from the decay of charmed mesons, a charm quark 
elastic scattering cross section much larger than that estimated from 
the perturbative QCD is required.
\end{abstract}

\section{INTRODUCTION}

Because of their large masses, charm quarks are produced very early 
and propagate through the quark-gluon plasma formed in relativistic 
heavy ion collisions. Any modifications of charm quark spectrum thus
carry information on the properties of the quark-gluon plasma. Although 
charmed hadrons are at present not directly observable in central 
nucleus-nucleus collisions at the Relativistic Heavy Ion Collider (RHIC), 
experimental data on the transverse momentum spectrum of electrons from 
their decays have already provided useful information on the interaction
of charm quarks in the quark-gluon plasma. For example, the transverse 
momentum spectrum of these electrons is found to be insensitive to 
the charm final-state interactions as results from both the PYTHIA 
model and the blastwave hydrodynamic model are consistent with the 
experimental data \cite{Batsouli:2002qf}. On the other hand, the 
large elliptic flow of these electrons is consistent with the 
prediction of the coalescence model \cite{Lin:2003jy,Greco:2003vf} 
which assumes that charm and light quarks are in thermal equilibrium 
and have same elliptic flow. In the present talk, we discuss 
in the framework of the AMPT model 
\cite{Zhang:1999bd,Lin:2000cx,Lin:2001yd,Lin:2001zk,Lin:2003ah,Lin:2004en}
the mechanism for the generation of charm quark elliptic flow and
the dependence of its value on the charm scattering cross section 
in the quark-gluon plasma \cite{Zhang:2005vq}. 
 
\section{THE AMPT MODEL}

The AMPT model has four components: initial conditions, parton cascade, 
hadronization, and hadron cascade. For studying charm elliptic flow, 
we use the version with string melting, in which hadrons that are
generated from the HIJING model \cite{Wang:1991ht} are converted to
partons according to their valence structures with a formation time that 
is determined by the transverse momentum of the parent hadron, in order 
to simulate the evolution of the energy stored in initial strings and 
the effects of particle production from the coherent color field. 
The space-time evolution of resulting partonic system is modeled 
by the ZPC model \cite{Zhang:1997ej}, which includes elastic 
scatterings between partons with a cross section given by the leading 
pQCD and regulated by a screening mass that is taken as a parameter 
to adjust the magnitude of the cross section. After the partonic system 
freezes out, closest quarks and anti-quarks are recombined into hadrons 
with their subsequent evolution simulated by the ART model 
\cite{Li:1995pr}. Using parton cross sections $6$-$10$ mb, the AMPT model 
with string melting can give a good description of measured low 
transverse momentum particle spectrum \cite{Chen:2004vh}, elliptic 
flow \cite{Lin:2001zk,Chen:2004vh}, higher-order anisotropic flows 
\cite{Chen:2004dv}, and the pion interferometry \cite{Lin:2002gc}. 

\section{SIMULATION OF CHARMED PARTICLES}

Charmed particles are rare particles even at RHIC as only about two pairs 
are produced in the mid-rapidity region of central Au+Au collisions 
at available top energies. To simulate charm particles efficiently, we 
use the perturbative method \cite{Randrup:1980qd} by introducing 
an enhancement factor for their production from initial 
hard scattering, so that each charmed particle carries a probability 
given by the inverse of corresponding enhancement factor, 
and neglects the effects due to charmed particle scattering 
on un-charmed particles. Using the power law parametrization of 
D meson spectrum measured in d+Au collisions by the STAR collaboration 
\cite{Adams:2004fc,Tai:2004bf,Ruan:2004if}, we first generate 
the transverse momentum distribution of D mesons between 
rapidity of -2 and +2 with their distribution in the transverse 
plane according to the positions of initial nucleon-nucleon collisions.
The initial phase-space distribution of charm quarks is then obtained 
by dissociating D mesons into their valence quarks after a formation 
time given by inverse of the D meson transverse momentum. The charm 
scattering cross section with other partons in the quark-gluon plasma 
is taken to be the same as the cross section for collisions between 
light quarks. Both cross sections of 3 mb, which is similar to that
given by the perturbative QCD, and 10 mb that is needed for describing
observables related to light quarks, are used in present study. 
At the freeze-out of partons, charm quarks are combined with light quarks into 
D mesons according to the coordinate-space coalescence model used in 
the AMPT model. For the scattering of charmed mesons with other hadrons,
their cross sections are simply taken to be the same as the charm-parton 
cross section as their effect on the charmed meson spectrum is 
insignificant. 

\section{RESULTS AND DISCUSSIONS}

From their transverse momentum spectra at mid-rapidity, charm quarks
are found to approach thermal equilibrium when their scattering cross 
sections increase. This result is reminiscent of the transition from the 
charm spectrum obtained from the PYTHIA to that of the blastwave 
hydrodynamic model. As in Ref.\cite{Batsouli:2002qf}, the spectrum of 
electrons from the decay of final charmed mesons is not very sensitive 
to charm final-state interactions and is compatible with experimental 
data \cite{Adler:2004ta} for both cross sections.

\begin{figure}
\center\includegraphics[width=100mm,height=100mm]{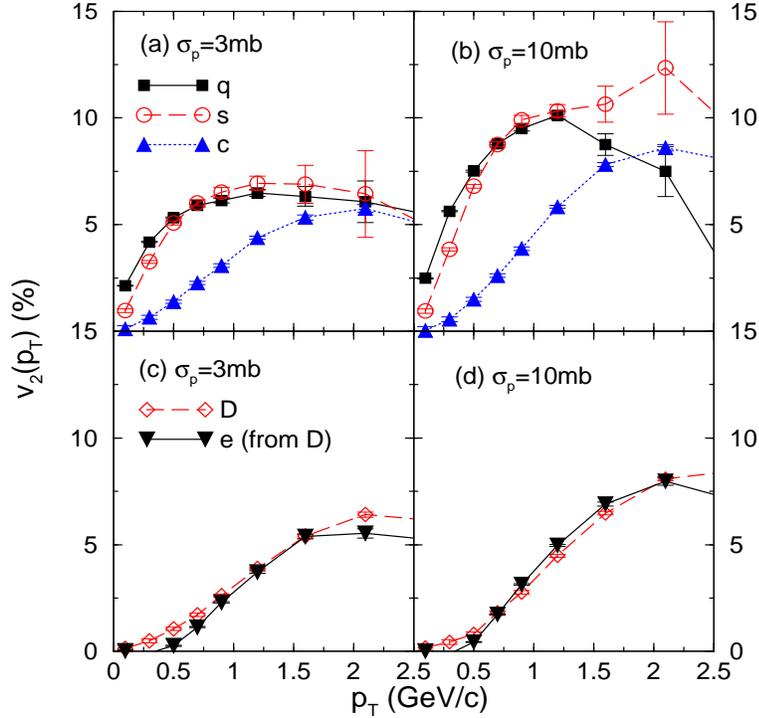}
\caption{Elliptic flows of quarks, D mesons, and electrons from D meson 
decay in minimum bias Au+Au collisions at $\sqrt{s_{NN}}=200$ GeV from 
the AMPT model.}
\label{fig:v2pt2}
\end{figure}

For elliptic flows, results from the AMPT model are shown in 
Fig.~\ref{fig:v2pt2}. It is seen from panels (a) and (b) that
the magnitude of charm quark elliptic flow increases with 
increasing parton scattering cross section.  For both parton cross 
sections, there is, however, a strong mass ordering of quark 
elliptic flows with charm quark elliptic flow saturating to about the 
same value at a larger transverse momentum compared with that 
of light quarks. The elliptic flows of D mesons and their decay
electrons are shown in panels (c) and (d). Both are seen to follow
essentially that of charm quarks as the momentum of a charmed 
meson is largely given by that of charm quark when light quarks 
have only bare masses as in the AMPT model. A larger charmed meson
elliptic flow of about 10\% at $p_t=2$ GeV/c is obtained
if an isotropic instead of screened Coulomb cross section is used 
for charm quark scattering as suggested recently in 
Ref.\cite{vanHees:2004gq} that charm quark scattering is dominated 
by the formation of charmed meson resonances in the quark-gluon 
plasma. To explain the large charm elliptic flow of more than 10\% 
in the available data \cite{Adler:2005ab,Laue:2004tf} thus requires
effects beyond the perturbative QCD in the quark-gluon plasma.

We have not included in the present study electrons from the decay 
of mesons consisting of bottom quark. Their contribution is expected 
to become important at $p_t>3$ GeV/c. Also, the effect of radiative energy 
loss on charm quark elliptic flow, which becomes non-negligible as the charm 
quark transverse momentum increases, is not considered.
For charm quarks with high transverse momentum, charm meson production 
will be mainly from fragmentation instead of recombination.
These effects need to be included for a more complete study of 
heavy quark elliptic flow at high transverse momentum. Our results 
are, nevertheless, consistent with other recent studies of charm 
collective flow 
\cite{vanHees:2004gq,Chen:2004cx,Molnar:2004ph,Bratkovskaya:2004ec,Moore:2004tg,Djordjevic:2005db,Petreczky:2005nh,vanHees:2005wb}, i.e., 
charm elliptic flow is sensitive to charm final-state interactions
and a large elliptic flow seen in available data requires a 
charm scattering cross section larger than that given by the 
perturbative QCD.

\end{document}